%% file: main.tex
\documentclass[journal=nalefd,manuscript=letter,layout=twocolumn,email=false]{achemso}
\let\oldmaketitle\maketitle
\let\maketitle\relax

\usepackage{graphicx} 
\usepackage{svg}
\usepackage{amsmath,amssymb}
\usepackage{color}
\usepackage{braket}
\usepackage{comment}

\title{Intrinsically slow cooling of hot electrons in CdSe nanocrystals compared to CdS}
\author{Matthew J. Coley-O'Rourke}
\affiliation{Department of Chemistry, University of California, Berkeley, California, 94720, United States}
\author{Bokang Hou}
\affiliation{Department of Chemistry, University of California, Berkeley, California, 94720, United States}
\alsoaffiliation{Materials Sciences Division, Lawrence Berkeley National Laboratory, Berkeley, California, 94720, United States}
\author{Skylar J. Sherman}
\affiliation{Department of Chemistry, University of Colorado Boulder, Boulder, Colorado 80309, United States}
\author{Gordana Dukovic}
\affiliation{Department of Chemistry, University of Colorado Boulder, Boulder, Colorado 80309, United States}
\alsoaffiliation{Department of Chemistry and Renewable and Sustainable Energy Institute (RASEI), University of Colorado Boulder, Boulder, Colorado 80309, United States}
\alsoaffiliation{Materials Science and Engineering, University of Colorado Boulder, Boulder, Colorado 80303, United States}
\author{Eran Rabani}
\affiliation{Department of Chemistry, University of California, Berkeley, California, 94720, United States}
\alsoaffiliation{Materials Sciences Division, Lawrence Berkeley National Laboratory, Berkeley, California, 94720, United States}
\alsoaffiliation{The Raymond and Beverly Sackler Center of Computational Molecular and Materials Science, Tel Aviv University, Tel Aviv, 69978, Israel}

\begin{document}
\twocolumn[
\begin{@twocolumnfalse}
\oldmaketitle
\begin{abstract}
\noindent
The utilization of excited charge carriers in semiconductor nanocrystals (NCs) for optoelectronic technologies has been a long-standing goal in the field of nanoscience. Experimental efforts to extend the lifetime of excited carriers have therefore been a principal focus. To understand the limits of these lifetimes, in this work, we theoretically study the timescales of pure electron relaxation in negatively charged NCs composed of two prototypical materials: CdSe and CdS. We find that hot electrons in CdSe have lifetimes that are five to six orders of magnitude longer than in CdS when the relaxation is governed only by the intrinsic properties of the materials. Although these two materials are known to have somewhat different electronic structure, we elucidate how this enormous difference in lifetimes arises from relatively small quantitative differences in electronic energy gaps, phonon frequencies, and their couplings. We also highlight that the effect of the phonons on electron relaxation is dominated by a Fr\"ohlich-type mechanism in these charged systems with no holes.
\end{abstract}
\end{@twocolumnfalse}
]

Semiconducting nanocrystals (NCs) are widely used in a variety of technologies for the conversion between optical and electronic energy.\cite{boles2016self,pietryga2016spectroscopic,almutlaq2024engineering,montanarella2022three} One of their promising applications is in the emission and manipulation of light within the near infrared and terahertz frequency range, where NCs can surpass the performance of bulk materials.\cite{jeong2016mid,lhuillier2017recent,livache2019colloidal,asgari2021quantum,montanarella2022three} However, generating low-energy light in devices presents several challenges, particularly the need for materials that support spontaneous emission across narrow energy gaps required for infrared light. This often leads to long radiative lifetimes, increased nonradiative decay, and consequently low quantum yields.\cite{Guyot-Sionnest2005,pandey2008slow} 

\begin{figure}[t!]
    \centering
    \includegraphics[width=0.8\linewidth]{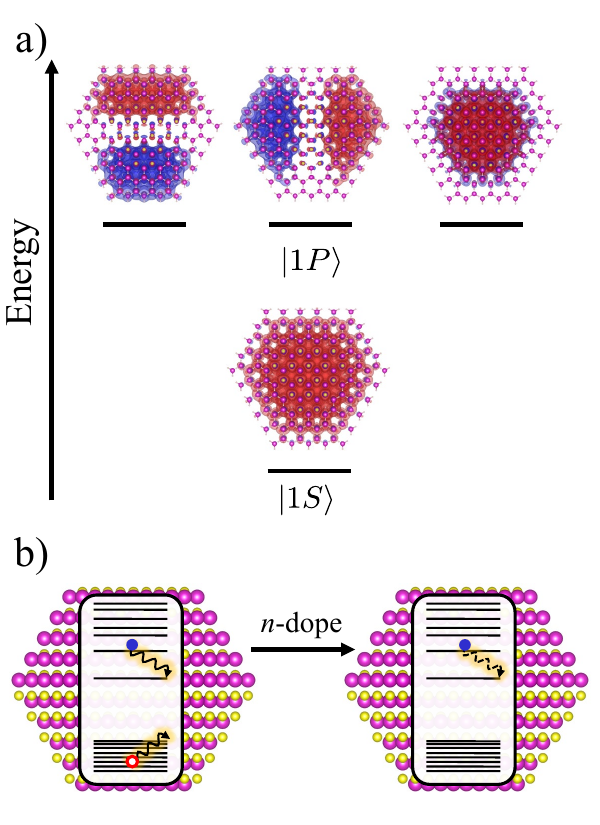}
    \caption{Overview of the hot electron relaxation in a \textit{n}-doped NC. (a) Isosurface plots of the $\ket{1P}$ and $\ket{1S}$ electron states in a CdS NC, calculated using the semiempirical pseudopotential model. The positive and negative phases of the electronic wave functions are shown in red and blue, respectively. (b) Comparison of the nonradiative relaxation of an exciton in a neutral NC and a hot electron in a doped NC. The electron and hole are represented by blue and red circles, and dissipation through phonons is illustrated with curly yellow arrows. This work is focused on the decay of hot electrons without a hole.}
    \label{fig1}
\end{figure}

In bulk materials, the continuous density of phonons and charge carriers facilitates rapid nonradiative cooling, leading to a loss of emission. In contrast, NCs exhibit discrete energy levels for electrons, holes, and phonons due to the quantum confinement effect,\cite{efros1982interband, brus1983simple} which can greatly reduce nonradiative processes by minimizing energy-conserving scattering events.\cite{klimov2014multicarrier,melnychuk2021multicarrier,jasrasaria2023circumventing} Notably, the transition between the $\ket{1P}$ and $\ket{1S}$ electron states in II-VI semiconductor NCs has been proposed as an optimal system for generating infrared and terahertz photons (see Fig.~\ref{fig1}).\cite{pandey2008slow,jeong2016mid,samadi2020semiconductor,kamath2021toward}

This prospect has motivated many studies of hot electron cooling in II-VI semiconductor NCs.\cite{Guyot-Sionnest2005,pandey2008slow,tisdale2010hot,diroll2020intersubband,jiang2011hot,sippel2013two,liu2014uncovering,li2017slow,wang2021spin,shulenberger2021photocharging,oriel2024intraband,wang2024hot}  Experimentally, creating and measuring hot electrons with long lifetimes in NCs is challenging. When the ground state is excited optically, it creates an electron-hole bound pair (exciton), resulting in rapid nonradiative relaxation of both charge carriers~\cite{nozik2021quantization} to their corresponding band edges due to an exchange-assisted process.\cite{efros1995breaking,kharchenko1996auger} To address this, many studies have aimed at spatially separating the electron and hole or eliminating the hole entirely.\cite{pandey2008slow,maity2015slow} In principle, removing the hole entirely creates a negatively charged (or $n$-doped) NC,\cite{shulenberger2021photocharging} and if this electron could be made hot, then it cannot cool via the exchange-assisted mechanisms, but eventually relaxes to the $\ket{1S}$ ground state by interacting with the NC lattice vibrations on longer timescales (see Fig.~\ref{fig1}). The challenge for low-frequency light generation is to identify conditions where this nonradiative phonon-assisted relaxation from the $\ket{1P}$ to the $\ket{1S}$ state is slower than the radiative lifetime of the $\ket{1P}$ electron, estimated to be in the sub-microsecond timescale~\cite{Guyot-Sionnest2005} (see also Fig.~\ref{fig:rates} below).

In this Letter, we investigate hot electron cooling in CdSe and CdS NCs under conditions where the transition rate between electron states is governed solely by phonon scattering, known as the ``phonon bottleneck" regime. We first present a model developed within the semiempirical pseudopotential framework, which incorporates polar bonding and Fröhlich interactions between electrons and nuclei. Using this model, we analyze the phonon-limited cooling rates from the $\ket{1P}$ manifold to the $\ket{1S}$ state (see Fig.~\ref{fig1}), demonstrating a strong dependence on the composition and size of the NCs. We find that in CdS, the lifetimes vary by three orders of magnitude across a size range of $3.7$ to $6.2$ nm in diameter. Moreover, we demonstrate that lifetimes in CdSe are $5-6$ orders of magnitude longer than those in similarly sized CdS NCs across the size range studied, and exceed recent experimental measurements. We then explore how minor quantitative differences between CdS and CdSe contribute to this substantial variation in the ``phonon bottleneck" regime of hot electron cooling rates. Finally, we examine the cooling of higher energy electron states, revealing that the $P \rightarrow S$ transition is the rate-limiting step.

\begin{figure*}[ht]
    \centering
    \includegraphics[width=0.9\linewidth]{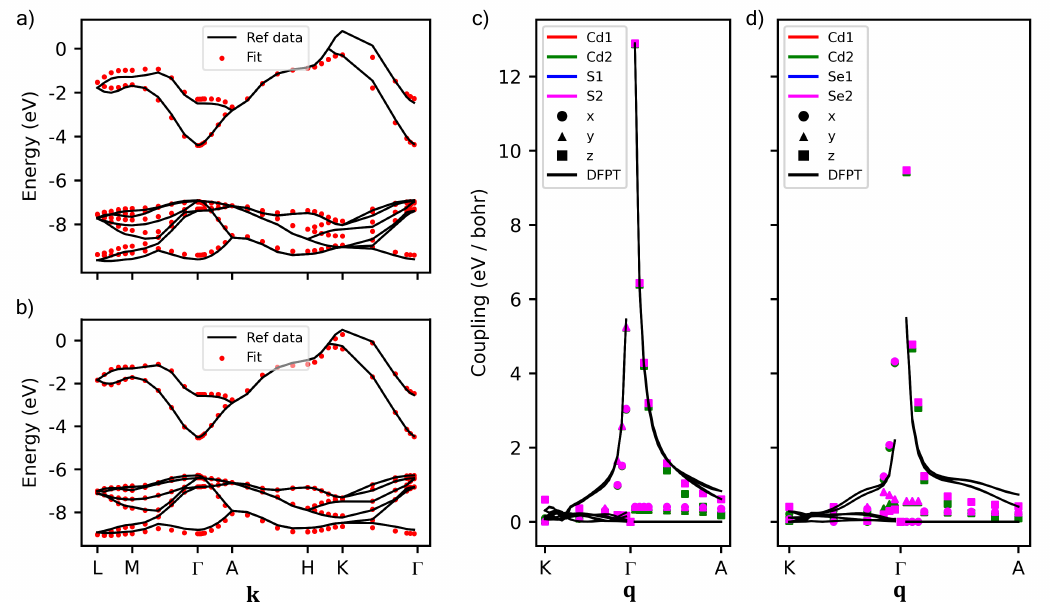}
    \caption{Overview of the model and fit fidelity. Band structures of bulk CdS (a) and CdSe (b), computed from the pseudopotential Hamiltonian and compared to reference data.\cite{cohen1967} A slice of the bulk CdS (c) and CdSe (d) electron-phonon coupling tensor $g_{i,j}^{\alpha}(\mathbf{k},\mathbf{q})$, where $\mathbf{q}$ is the phonon wavevector (see SI for more details), and we plot results for $\mathbf{k} = \Gamma$, $i = j =$ conduction band. We plot all modes $\alpha$ in terms of their atomic positions and Cartesian axes. Reference values are computed from density functional perturbation theory (DFPT).\cite{gonze1992,giustino2017electron}
    }
    \label{fig:fitting}
\end{figure*}

To describe the hot electron cooling process, we simulate the quantum dynamics of a model Hamiltonian consisting of three terms,\cite{hou2023incoherent,jasrasaria2023circumventing}
\begin{align}
    &\hat{H} = \hat{H}_{el} + \hat{H}_{ph} + \hat{H}_{el-ph} = \sum_i \varepsilon_i \ket{\phi_i} \bra{\phi_i} + \nonumber \\
    &\sum_{\alpha} \left [ \frac{\hat{P}^2_{\alpha}}{2} + \frac{\omega^2_{\alpha}}{2} \hat{Q}^2_{\alpha} \right] + \sum_{\alpha, i,j} V_{ij}^{\alpha} \ket{\phi_i}\bra{\phi_j} \hat{Q}_{\alpha}
    \label{eqn:model_ham}
\end{align}
The first term describes the electrons in the nanocrystal in the basis of the corresponding one-electron eigenfunctions $\ket{\phi_i}$ and associated energies $\varepsilon_i$ (further details below). The second term describes the NC nuclear motions in the normal mode representation, each with its associated (mass-weighted) momentum $\hat{P}_{\alpha}^2$ and displacement $\hat{Q}_{\alpha}^2$. The normal modes and the corresponding frequencies $\omega_\alpha$ were computed by diagonalizing the dynamical matrix using the Stillinger-Weber force field at the equilibrium NC geometry.\cite{zhou2013stillinger} The final term describes the coupling between electron states and the normal modes to first order in the normal modes displacements, $Q_\alpha$, with magnitudes given by~\cite{jasrasaria2021interplay} $V_{ij}^{\alpha} = \bra{\phi_j} \frac{\partial V_{\rm elec}}{\partial Q_{\alpha}} \ket{\phi_i}$. 

\begin{figure*}[t!]
    \centering
    \includegraphics[width=0.9\linewidth]{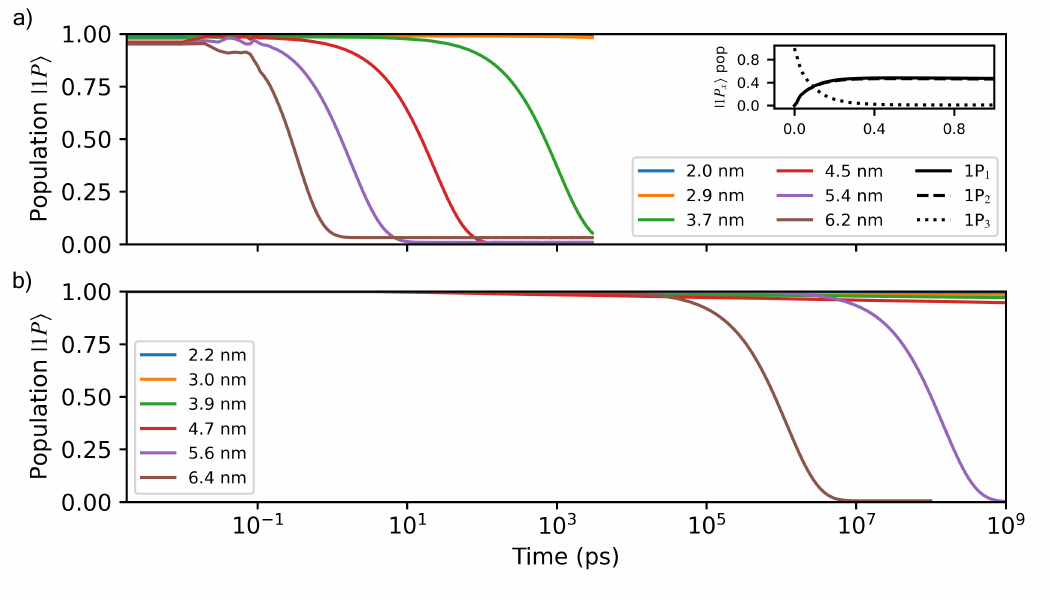}
    \caption{Real time dynamics of electron population in the $\ket{1P}$ manifold for various diameters of (a) CdS and (b) CdSe NCs ($T=300K$). The inset shows a representative example of the sub-picosecond mixing between the 3 states in the $\ket{1P}$ manifold (4.5 nm CdS is shown). When the $\ket{1P}$ population decays to zero, the electron is fully relaxed to the $\ket{1S}$ state.}
    \label{fig:real_time}
\end{figure*}

The electron states and $V_{ij}^{\alpha}$ were computed from a semiempirical pseudopotential model.\cite{wang1995local,fu1997local,zunger1997pseudopotential,rabani1999electronic,jasrasaria2022simulations} Two important aspects of our model differ from previous work. First, our model is parameterized to recover most accurately the states and energies of \textit{excess} electrons above the bandgap, since we are focused on electron dynamics in negatively charged NCs. Second, our model explicitly captures long range Fr{\"o}hlich-type interactions between the electrons and nuclei,\cite{frohlich1950xx} which have a large impact on the dynamics in charged systems. The total potential felt by an electron ($V_{\rm elec}$) was modeled as a sum of spherical, atom-centered functions which include the well-established short range functions~\cite{wang1994electronic} $v_{\rm short}$ plus an additional long-range Coulomb tail:
\begin{align}
    V_{\rm elec}(\mathbf{r}) = \sum_{\alpha}^{N_{\rm atom}} &\Big[ v_{\rm short}^{\alpha}(|\mathbf{r}-\mathbf{R_{\alpha}}|) \Big.\nonumber \\
    &\Big. + c^{\alpha} \frac{\mathrm{erf}(\lambda \: |\mathbf{r}-\mathbf{R_{\alpha}}|)}{|\mathbf{r} - \mathbf{R_{\alpha}}|} \Big].
\end{align}
Here $\mathbf{r}$ labels the position of the electron and $\mathbf{R}_{\alpha}$ the position of nucleus $\alpha$ in the NC. The parameters used in $v_{\rm short}^{\alpha}(|\mathbf{r}|)$ as well as $\lambda$ and $c^{\alpha}$ were fitted to reproduce the bulk band structure and electron-phonon coupling determined from first-principles calculations (see Fig.~\ref{fig:fitting}). The addition of the Coulomb tail into the atom-centered potentials allows us to capture the effects of the variable effective charge on atoms that form polar bonds such as Cd-S and Cd-Se, giving rise to Fr{\"o}hlich-type interactions. This is reflected in Fig.~\ref{fig:fitting}c,d, where our model reproduces the \textit{ab initio} divergence of electron-phonon coupling strength as $\mathbf{q} \rightarrow 0$ ($\Gamma$ point). See the the SI for more details on the pseudopotential model and the parameterization of the Hamiltonian.

The real-time dynamics of hot electrons (with no hole counterpart) transitioning from the $\ket{1P}$ manifold to the $\ket{1S}$ state were simulated using a quantum master equation that describes populations and memory effects to second order in the dressed electron-phonon couplings,\cite{nitzan2024chemical,jasrasaria2023circumventing,izmaylov2011nonequilibrium} based on the model Hamiltonian (cf. Eq.~\eqref{eqn:model_ham}). To account for higher-order multiphonon relaxation channels in our perturbative description of the dynamics, we performed a unitary polaron transformation~\cite{nitzan2024chemical} on the Hamiltonian given by Eq.~\eqref{eqn:model_ham} (see SI for more details).

The population dynamics for a series of CdSe and CdS NCs of different sizes are shown in Fig.~\ref{fig:real_time}. These dynamics describe pure nonradiative relaxation of the charged NCs from the $\ket{1P}$ manifold to the $\ket{1S}$ state at $300$~K. Our model excludes trap states in this energy gap and the relaxation proceeds entirely via coupling to lattice vibrations. One prominent feature of the results is the rapid decrease in $\ket{1P}$ lifetime with increasing particle size in CdS. The lifetime decreases by $\sim 3$ orders of magnitude as the particle diameter increases from $3.7$~nm to $6.2$~nm. Another prominent feature of the dynamics in Fig.~\ref{fig:real_time} is the significant slowdown of relaxation in CdSe compared to CdS. The relaxation dynamics are so slow in CdSe (more below) that we were only able to converge the dynamics for the largest particles. Comparing the results for CdS and CdSe for the largest particles, we find the CdSe dynamics are slower by approximately six orders of magnitude than in similarly sized CdS NCs. In addition, the short-time dynamics ($t\le 100 fs$) in CdS exhibit a damped oscillation, assigned to the coupling to optical vibrational modes, not observed in CdSe due to the weaker electron-phonon couplings and the longer hot electron lifetimes.

\begin{figure*}[t!]
    \centering
    \includegraphics[width=0.9\linewidth]{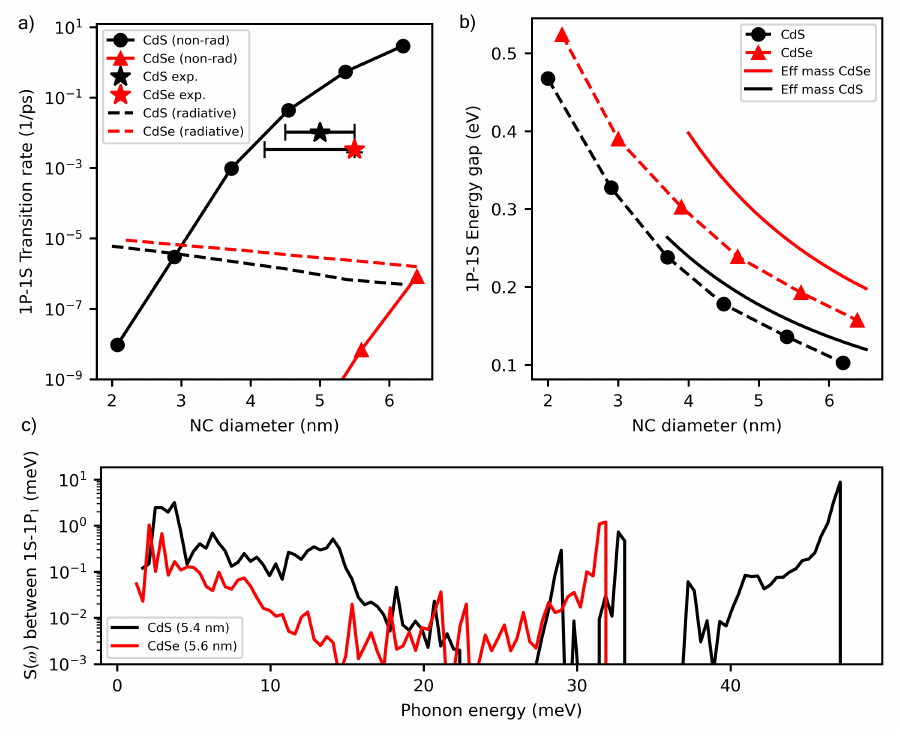}
    \caption{(a) 1P-1S non-radiative cooling rates for CdS and CdSe, corresponding to the dynamics from Fig.~\ref{fig:real_time}, along with the radiative emission rates calculated using the pseudopotential model (see SI for more details). We also show experimental values for CdS~\cite{skylar2024} and CdSe.\cite{guyot2005intraband} The major contributing factors to the non-radiative rates are (b) the 1P-1S energy gaps, and (c) the electron-phonon coupling displayed in terms of the spectral density $S_{\ket{P},\ket{S}}(\omega)$, shown here for a representative NC size $\approx 5.5$nm. In (b) the dashed lines result from our pseudopotential model, while the solid lines are determined by an effective mass model for comparison.\cite{bawendi1996,korolev2015} For visualization in (c), we plot the $\delta$-function in Eq.~\eqref{eq:Sw} as a box function with width $dE=0.4$ meV and height $1/dE$.
    }
    \label{fig:rates}
\end{figure*}

These dramatic differences in the behavior of the electron in CdS and CdSe can be seen even more clearly in Fig.~\ref{fig:rates}a, where we plot the relaxation rates computed from fitting the population dynamics to an exponential decay. To explain the discrepancy, we plot in Fig.~\ref{fig:rates}b the computed energy gap between the $\ket{1P}$ manifold and the $\ket{1S}$ state as a function of NC size. For the $P\rightarrow S$ transition to occur in our model, the $\ket{1P}$ electron needs to transfer this amount of energy to the lattice vibrations. To characterize how much each normal mode can contribute to this process, we plot in Fig.~\ref{fig:rates}c the spectral densities, defined as,
\begin{equation}
    S_{ij}(\omega) = \pi \sum_{\alpha} \omega_{\alpha} \lambda^{\alpha}_{ij} \delta(\omega - \omega_{\alpha}).
    \label{eq:Sw}
\end{equation}
Here, $\lambda^{\alpha}_{ij} = \frac{1}{2} (\frac{V_{ij}^{\alpha}}{\omega_{\alpha}})^2$ is the mode-wise reorganization energy and $\delta(\omega)$ is the Dirac delta function. In both CdS and CdSe, the high frequency optical modes ($33$~meV for CdSe, $50$~meV for CdS) have the largest contributions via the Fr\"ohlich interaction, with other significant peaks coming predominantly from low frequency acoustic modes. However, throughout nearly the entire frequency range the CdS modes have peaks that are $\sim 5-10$ times larger than in CdSe. The combined effect of all modes can be summarized by the total reorganization energy for the transition, $\lambda_{P,S} = \pi^{-1} \int d\omega S_{P,S}(\omega)/\omega$, which we compute to be $\lambda_{P,S} \approx 14$~meV in CdS and $\lambda_{P,S} \approx 2.4$~meV in CdSe. Therefore, based on the magnitude of the reorganization energy, $\lambda_{P,S}$, which reflects the strength of the electron-phonon coupling, we can conclude that CdSe will exhibit slower relaxation than CdS. However, other factors, which we will discuss next, play a more significant role in influencing the relaxation dynamics.

In both NCs, the highest optical frequencies are smaller than the $1P-1S$ energy gaps (Fig.~\ref{fig:rates}b,c) and consequently, multi-phonon processes are required to enable the transition. In this regime, the transition rate is expected to depend exponentially on the $1P-1S$ energy gaps,\cite{Nitzan75} and since this gap depends on the NC size, we expect to the nonradiative rate to increase rapidly with the diameter of the NC.  Additionally,  we expect the rate to contain a multiplicative factor which decays exponentially with the number of phonons needed to achieve resonance $\sim \frac{E_P - E_S}{E_{ph}}$.\cite{egorov1997vibrational} These factors explains our above observation of a rapid decrease in $\ket{1P}$ lifetime with increasing CdS NC size; within this size range, the energy gap decreases significantly, dropping by more than 50\% as the NC size increases from $3.7$~nm to $6.2$~nm. 

Comparing the results for CdSe and CdS across various NC sizes, we observe that the $1P-1S$ gap in CdSe is consistently $50-60$~meV larger than that of similarly sized CdS. By fitting the CdS rates in Fig.~\ref{fig:rates}a to an exponential function of the energy gap, we can estimate that this difference alone contributes only about 1 order of magnitude to the total difference between CdS and CdSe (for intermediate to large sizes). The additional discrepancy emerges because CdSe has lower energy optical modes, and the combination of larger gaps and lower frequencies requires higher order multiphonon processes to conserve energy during the transition. For example, for an intermediate-sized NC ($4.5$~nm), CdSe would need approximately $10$ optical phonons to bridge the gap, whereas CdS would only require about $4$ optical phonons. This factor together with the smaller electron-phonon couplings in CdSe contributes the additional multiplicative factor~\cite{egorov1997vibrational} to the nonradiative decay rate that is $\sim 5$ orders of magnitude smaller in CdSe compared to CdS.

The discrepancies between our calculations and the experimental measurements for CdSe suggest that additional factors not considered in our model may play a significant role in explaining the channels for hot electron cooling in experiments. In particular, our model makes several assumptions about the surface of the NCs, and a more accurate treatment may be necessary to account for interactions with ligand vibrations,\cite{guyot2005intraband} trap states, or surface charge residues (holes or other electrons). While these factors could qualitatively alter the cooling mechanism, our results suggest that by reducing NC size and minimizing trap states near the $S-P$ gap, it is possible to achieve sufficiently long cooling times for infrared and terahertz technologies.

\begin{figure}[t!]
    \centering
    \includegraphics[width=0.9\linewidth]{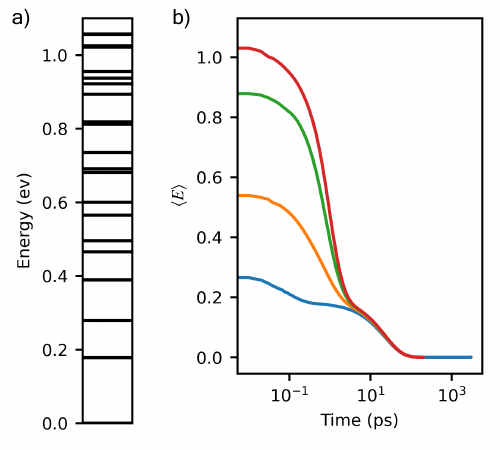}
    \caption{(a) Energy spectrum of electron states higher than $\ket{1S}$ in 4.5nm CdS. The energy of $\ket{1S}$ is set to 0. (b) Dynamics of hot electrons starting from various states above the $\ket{1P}$ manifold (each in a different color). The y-axis shows the energy expectation value of the electron as its population spreads out over multiple states during its time evolution. The cooling timescales to $\ket{1S}$ are all very similar to the $1P \rightarrow 1S$ transition (see Fig.~\ref{fig:real_time})}
    \label{fig:high-energy}
\end{figure}

As a final point of analysis, we note that experimental preparation of hot electrons without a hole counterpart is not straightforward~\cite{wang2021spin} and therefore may not be able to easily ``initialize" the electron in the $\ket{1P}$ manifold, as we have done in Figs.~\ref{fig:real_time} and \ref{fig:rates}. For example, in schemes that rely on trion recombination to generate a hot electron,\cite{skylar2024} the electron would start with energy comparable to the optical gap above the band edge. In Fig.~\ref{fig:high-energy}, we analyze the dynamics of electrons that begin in states higher in energy than $\ket{1P}$. Fig.~\ref{fig:high-energy}a shows the energy spectrum of electron states above $\ket{1S}$ in a representative example of 4.5 nm CdS. 

Beyond the lowest-lying $1P\rightarrow 1S$ gap, there are other gaps that appear at higher energies that are comparable in size to the $1P-1S$ gap. It would be reasonable to expect that these sizable gaps contribute additional ``bottlenecks'' in the hot electron cooling process. However, we find that in general this is not the case. In Fig.~\ref{fig:high-energy}b we show the real time dynamics of a hot electron in a $4.5$~nm CdS NC. In this case, the higher energy electron is able to decay to the $\ket{1P}$ manifold faster than the $1P\rightarrow 1S$ transition occurs because all of the higher energy gaps are still small enough to allow for rapid relaxation to the $\ket{1P}$ manifold. Due to the exponential dependence of the rate on the gap, even the high energy gaps that are only $\sim 40$meV smaller than the $1P-1S$ gap still decay much more rapidly. We find that the $1P-1S$ gap is the largest across both materials and all NC sizes studied, suggesting that the rates measured in experiments that start with very hot electrons are still dominated by the $P \rightarrow S$ transition.    

In summary, we have used a model Hamiltonian parameterized by a semiempirical pseudopotential method to simulate the nonradiative relaxation of hot electrons with no counterpart hole in CdS and CdSe NCs. When the dynamics are governed solely by intrinsic electron-phonon coupling, we find that electrons in the $\ket{1P}$ state decay to the $\ket{1S}$ state $\approx 5-6$ orders of magnitude slower in CdSe than CdS. Additionally, we find that the rate changes rapidly with system size, increasing by three orders of magnitude as the NC size is increased. The differences between CdS and CdSe NCs, along with the size dependence of the nonradiative rate, were explained in terms of the coupling strength, the order of the multiphonon process, and the energy gaps between the $\ket{1P}$ and $\ket{1S}$ states. While our calculations of the nonradiative lifetimes for CdS agree with experiments,\cite{skylar2024} they disagree with recent experiments for CdSe that found no change in the $\ket{1P}$ lifetime as the size was varied from $4.2$~nm diameter to $5.5$~nm.\cite{wang2021spin} While the experimental trends with size are still not clear,\cite{Guyot-Sionnest2005} the lifetimes measured in these experiments are indeed substantially shorter than is predicted by our simulations and require further work to fully understand the dominant experimental factors causing hot electron relaxation in CdSe.

\section*{Acknowledgments}
This work was primarily supported by the U.S. Department of Energy, Office of Science, Office of Advanced Scientific Computing Research and Office of Basic Energy Sciences, Scientific Discovery through Advanced Computing (SciDAC) program under Award No. DE-SC0022088. In addition, the quantum dynamical simulations were supported by the U.S. Department of Energy, Office of Science, Office of Basic Energy Sciences, Materials Sciences and Engineering Division, under Contract No. DEAC02-05-CH11231 within the Fundamentals of Semiconductor Nanowire Program (KCPY23). Computational resources were provided in part by the National Energy Research Scientific Computing Center (NERSC), a U.S. Department of Energy Office of Science User Facility operated under Contract No. DEAC02-05CH11231. The experimental portion of this work was supported primarily by the Air Force Office of Scientific Research under AFOSR Award Nos. FA9550-19-1-0083 and FA9550-22-1-0347. S.J.S. acknowledges support from the Edward L. King and Marion L. Sharrah fellowships at University of Colorado Boulder.
\input{main.bbl}

\end{document}


\maketitle

\section{Semi-empirical pseudopotentials with long-range effects}
\label{sec:pp}
For the computation of the electronic structure of the CdS quantum dot, we employed the semi-empirical pseudopotential method.\cite{zunger1997pseudopotential}  This approach has been used to describe the vibronic and optical properties of nanocrystals (NCs) and has been validated for III-V,\cite{gupta2023composition} II-VI,\cite{philbin2018electron,jasrasaria2022simulations,lin2023theory,brosseau2023ultrafast} and perovskite NCs.\cite{weinberg2023size} In this approach, the total potential felt by an electron near the band edge is decomposed as a sum of atom-centered, spherically symmetric functions,
\begin{equation}
    V_{\rm tot}(\vecr) = \sum_{\mu} v_{\mu}(|\vecr-\vecR|),
\end{equation}
where $\vecr$ is the position of the electron and $\vecR$ is the position of atom $\mu$.
In previous works, the atomic potentials have been parameterized by short-range forms in real space, i.e. $v_{\mu}(|\vecr-\vecR|) \sim \exp(-|\vecr-\vecR|^2)$ for $|\vecr-\vecR| \rightarrow \infty$. As noted above, this choice has been satisfactory to accurately describe optical excitations in a variety of systems. In part, this choice has been made due to the relative ease of transforming the potentials between real space and reciprocal space. This is a crucial component of the method, in which the functional forms are accurately parameterized by fitting to \textit{bulk} band structures, and then transformed to real space for simulation of the NC.

\begin{figure}[t!]
    \centering
    \includegraphics[width=0.9\linewidth]{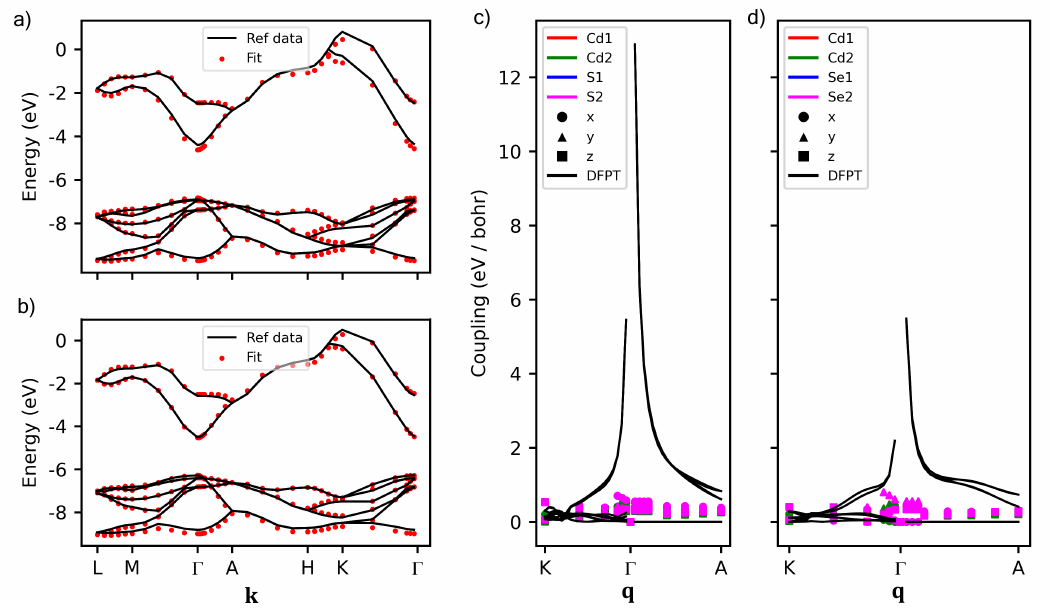}
    \caption{Failure of short-range pseudopotentials to capture Fr\"ohlich coupling in CdSe and CdS. Using short-range pseudopotentials~\cite{jasrasaria2022simulations}, the band energies of (a) CdS and (b) CdSe can be reproduced well. However, the electron coupling to atomic motions cannot recover \textit{ab initio} DFPT results in either (c) CdS or (d) CdSe. Plotted in (c)-(d) is a slice of the bulk electron-phonon coupling tensor $g_{i,j}^{\mu}(\mathbf{k},\mathbf{q})$, (see Eq.~\eqref{eq:bulk_cpl}), with $\mathbf{k} = \Gamma$, $i = j =$ conduction band. We plot all modes $\mu$ in terms of their atomic positions and Cartesian axes.}
    \label{fig:bad_cpl}
\end{figure}

The consideration of \textit{charged} excitations in this work, rather than neutral optical excitations, has revealed limitations of the traditional pseudopotential form. While the energies are still well described, the couplings between electrons and nuclear motion are not (see Fig~\ref{fig:bad_cpl}). To assess the pseudopotential and eventually parameterize it to recover accurate couplings, we computed ab initio reference data in bulk CdS and CdSe using density functional pertubation theory (DFPT)~\cite{gonze1992,giustino2017electron} with the PBEsol functional in the Quantum Espresso software package~\cite{giannozzi2017advanced}. The bulk couplings are expressed,
\begin{equation}
    g_{ij}^{\mu}(\mathbf{k}, \mathbf{q}) = \bra{\psi_{i}(\mathbf{k})} \frac{\partial \tilde{v}_{\mu}(|\vecg_i - \vecg_j|; \vecR)}{\partial \vecR(\mathbf{q})} \ket{\psi_{i}(\mathbf{k}+\mathbf{q})},
    \label{eq:bulk_cpl}
\end{equation}
where $\mathbf{k}$ is the electronic wavevector, $\mathbf{q}$ is the atomic displacement (phonon) wavevector, $\psi_{i,j}$ are one-electron eigenfunctions in bands $i$ and $j$, and $\tilde{v}_{\mu}$ is the Fourier transform of the atomic pseudopotential to reciprocal space, where $\vecg_i$ is a plane wave basis vector.

The most prominent feature of the ab initio couplings in both CdS and CdSe is the divergence as $\mathbf{q} \rightarrow 0$ ($\Gamma$-point). This is a signature of the Fr\"ohlich coupling mechanism~\cite{frohlich1950xx}, which emerges from the long-range interactions between electrons and nuclei with an effective charge. We note that this effect is present in \textit{all} semiconductors that form polar bonds. When considering optical excitations, this effect may be small due to the equal and opposite charge of the bound electron and hole, but in charged systems it is dominant.


In this work, we account for the long range Fr\"ohlich-type interactions by adding a long-range Coulomb term to the well-established short range functions.\cite{wang1994electronic} The total potential is thus decomposed in reciprocal space as,
\begin{equation}
\begin{split}
    V_{\rm tot}(\vecg_i, \vecg_j) &= \sum_{\mu} \tilde{v}_{\mu}(|\vecg_i - \vecg_j|; \vecR) = \sum_{\mu} \left[ \tilde{v}_{\mu}^{\rm short}(|\vecg_i - \vecg_j|; \vecR) + \tilde{v}_{\mu}^{\rm long}(|\vecg_i - \vecg_j|; \vecR) \right] =\\ 
    &= \sum_{\mu} \left[ a_0^{\mu} \frac{ |\vecg_i - \vecg_j|^2 - a_1^{\mu}}{a_2^{\mu} \exp(a_3^{\mu} |\vecg_i - \vecg_j|^2) - 1} - 4 \pi a_{4}^{\mu} \frac{\exp(\frac{-|\vecg_i - \vecg_j|^2}{4\lambda^2})}{|\vecg_i - \vecg_j|^2} \right] \: e^{-i (\vecg_i-\vecg_j) \cdot \mathbf{R_{\mu}}}
\end{split}
\label{eq:pot}
\end{equation}
Here, the parameters $\{ a_0, a_1, a_2, a_3, a_4, \lambda \}$ are tuned to reproduce the bulk band structure~\cite{cohen1967} and ab initio electron-phonon coupling reference data (see Fig.~2 in the main text), as well as the conduction band effective masses reported from experiments~\cite{madelung2004semiconductors}. The first term in Eq.~\eqref{eq:pot} is a well established short-range function in real space.\cite{wang1994electronic,rabani1999electronic} We introduce the second term, which Fourier transforms to the long-range Coulomb potential $a_4 \frac{\mathrm{erf}(\lambda |\mathbf{r} - \mathbf{R}_{\mu}|)}{|\mathbf{r} - \mathbf{R}_{\mu}|}$ in real space, enabling our model to capture the Fr{\"o}hlich-type interaction (see Fig.~2 in the main text). Crucially, the $a_4^{\mu}$ parameters of the long-range term must satisfy the charge neutrality-type condition $\sum_{\mu \in {\rm unit\: cell}} a_4^{\mu} = 0$.

For our simulations, we use the parameter values listed in Tables~\ref{tab:coefficients_S}-\ref{tab:coefficients_Se}. The atom-centered potentials are then converted to a real-space form, denoted $v_{\mu}(|\mathbf{r} - \mathbf{R}_{\mu}|)$ by standard three-dimensional Fourier transform. These real-space potentials are more amenable to simulating nanosystems.

\begin{table}[h!]
\centering
\begin{tabular}{|c|c|c|c|c|c|c|}
\hline
 & \( a_0 \) & \( a_1 \) & \( a_2 \) & \( a_3 \) & \( a_4 \) & \( \lambda \) \\ 
\hline
Cd & -31.450808 & 1.665222 & -0.16198998 & 1.672854 & -0.625 & 0.2 \\ 
\hline
S & 7.665165 & 4.444229 & 1.384734 & 0.2584866 & 0.625 & 0.2 \\ 
\hline
\end{tabular}
\caption{Table of the CdS pseudopotential coefficients \( a_0 \) to \( a_4 \) and $\lambda$ for Cd and S}
\label{tab:coefficients_S}
\end{table}

\begin{table}[h!]
\centering
\begin{tabular}{|c|c|c|c|c|c|c|}
\hline
 & \( a_0 \) & \( a_1 \) & \( a_2 \) & \( a_3 \) & \( a_4 \) & \( \lambda \) \\ 
\hline
Cd & -31.45180 & 1.3890 & -0.04487776 & 1.659671 & -0.248 & 0.2 \\ 
\hline
Se & 8.49210 & 4.35130 & 1.3558211 & 0.3236369 & 0.248 & 0.2 \\ 
\hline
\end{tabular}
\caption{Table of the CdSe pseudopotential coefficients \( a_0 \) to \( a_4 \) and $\lambda$ for Cd and Se. The Cd potential is not the same as Table~\ref{tab:coefficients_S} because the atom has a different long-range effective charge in CdSe compared to CdS.}
\label{tab:coefficients_Se}
\end{table}

To remove surface trap states from the gap and cancel macroscopic electric fields generated by the faceted surfaces of the quantum dot, we attach fictitious ligand potentials to the dangling bonds at the surface. These potentials take the general form,
\begin{equation}
    v_{lig}(|\mathbf{r}-\mathbf{R}_0|) = c_0 \exp(\frac{-|\mathbf{r}-\mathbf{R}_0|^2}{c_1}) + c_2  \frac{|a_4|\mathrm{erf}(\lambda |\mathbf{r} - \mathbf{R}_{0}|)}{|\mathbf{r} - \mathbf{R}_{0}|}.
\end{equation}
We use 4 different ligand potentials, due to the shape of the charge distribution in the bare faceted quantum dot. The long-range components of these potentials were fit to account for the balancing of the charge of the entire quantum dot and ligands. The parameters of the $4$ ligands potentials are given in Table~\ref{tab:ligands}. P3 passivates the dangling bonds of Cd atoms on facets that are perpendicular to the $c$-axis. P4 passivates the dangling bonds of S/Se atoms on facets that are perpendicular to the $c$-axis. P1 passivates Cd atoms on all other facets, while P2 passivates S/Se atoms on all other facets. The potentials have the same form for CdS and CdSe.

\begin{table}[h!]
    \centering
    \begin{tabular}{|c|c|c|c|c|}
        \hline
         & $c_0$ & $c_1$ & $c_2$ & passivates \\
         \hline
         P1 & 0.64 & 2.2287033 & -0.25152663  & Cd \\
         P2 & -0.384 & 2.2287033 & 0.24206696 & S/Se \\
         P3 & 0.64 & 2.2287033 & -0.24132418 & Cd \\
         P4 & -0.384 & 2.2287033 & 0.2535023 & S/Se \\
         \hline
    \end{tabular}
    \caption{Parameters for the 4 distinct ligand potentials on the surface of the CdS quantum dot. P3 and P4 passivate atoms only on facets that are perpendicular to the $c$-axis, while P2 and P3 passivates atoms on all other facets.}
    \label{tab:ligands}
\end{table}

\section{Model Hamiltonian and hot electron relaxation dynamics}
We utilized the semi-empirical pseudopotentials outlined above along with the Stillinger-Weber force field\cite{zhou_stillinger-weber_2013} to parameterize the electronic and nuclear Hamiltonian given by $H = H_{\rm el} + H_{\rm nu} + H_{\rm el-nu}$,\cite{jasrasaria2021interplay,hou2023incoherent} where 
\begin{align}
    &H_{\rm el} = \sum_{n} E_{n}\ket{\psi_n}\bra{\psi_n} \\
    &H_{\rm nu} = \sum_{\alpha}\frac{1}{2}{P}_{\alpha}^2 + \frac{1}{2}\omega_\alpha^2 Q_\alpha^2 \\
    &H_{\rm el-nu} =\sum_{n,m} \ket{\psi_n}\bra{\psi_m}\sum_{\alpha}V_{nm}^\alpha{Q}_{\alpha}.
\end{align}
For the electronic part, $E_n$ and $\ket{\psi_n}$ were obtained from the eigenvalue solution of quasi-electron Hamiltonian, $h_e$,
\begin{equation}
    h_e = \frac{1}{2}\nabla^2 + \sum_{\mu} v_{\mu}(|\mathbf{r} - \mathbf{R}_{\mu,0}|).
\end{equation}
We used the filter diagonalization technique~\cite{toledo2002very} to calculate the quasi-electron states near the bottom of the conduction band, including one $\ket{1S}$ and three $\ket{1P}$ electronic states. The nuclear degrees of freedom were represented using the Stillinger-Weber force field,\cite{zhou2013stillinger} with the reference geometry $\mathbf{R}_0$ being the energy minimum. The normal mode frequencies $\omega_\alpha$ and coordinates were calculated by diagonalizing the Hessian matrix at $\mathbf{R}_0$ under the harmonic approximation. $P_\alpha$ and $Q_\alpha$ denote the mass-weighted normal mode momentum and coordinates, respectively, for mode $\alpha$. The transformation between the atomic and normal modes is given by $Q_\alpha = \sum_{\mu,k}\sqrt{M_\mu}E_{\mu k, \alpha}(R_{\mu k}-R_{\mu k, 0})$, where $k=x, y, z$ and $M_\mu$ is the mass of atom $\mu$. The electron-phonon coupling between the and electron in state $\ket{\psi_n}$ and $\ket{\psi_m}$ with respect to mode $\alpha$ is denoted as $V_{nm}^\alpha$, and can be related to the derivative of the pseudopotential by~\cite{jasrasaria2021interplay}
\begin{align}
    V_{nm}^\alpha \equiv \bra{\psi_n}\frac{\partial h_{\rm e}}{\partial Q_\alpha}\ket{\psi_m} = \sum_{\mu, k} \frac{1}{\sqrt{M_\mu}}E_{\mu k, \alpha} \bra{\psi_n}\frac{\partial v_\mu}{\partial R_{\mu k}}\ket{\psi_m}. 
    \label{eq:coupling_B}
\end{align}

To calculate the transition rate $k_{n\to m}$ from state $\ket{\psi_n}$ to $\ket{\psi_m}$ (or 1P to 1S transition in this case), we first apply a unitary small polaron transformation, with displacement operator $e^S=e^{\sum_n S_n\ket{n}\bra{n}}$ and ${S_n} = {-\frac{i}{\hbar}\sum_{\alpha}\frac{V_{nn}^\alpha}{\omega_\alpha^2} P_\alpha}$, leading to a transformed Hamiltonian of the form:
\begin{equation}\label{eq:pt_ham}
    \Tilde{H} = e^S H e^{-S} = \sum_n\varepsilon_n\ket{\psi_n}\bra{\psi_m} +  \sum_\alpha \frac{P_\alpha^2}{2} + \frac{1}{2}\omega_\alpha^2Q_\alpha^2 + \sum_{nm}\ket{\psi_n}\bra{\psi_m}\sum_\alpha V_{nm}^\alpha e^{S_n}Q_\alpha e^{-S_m}
\end{equation}
In the above, the new system energy $\varepsilon_n = E_n - \lambda_n$ is shifted by the reorganization energy $\lambda_n = \sum_\alpha \frac{(V_{nn}^\alpha)^2}{2\omega_\alpha^2}$ relative to $E_n$. The purpose of performing the polaron transform is to renormalize the interstate couplings $V_{12}^\alpha$, improving the performance of perturbation theory while incorporating multi-phonon processes.

To compute the rate  $k_{n\to m}$, we employ Fermi's golden rule treating the dressed off-diagonal electron-phonon coupling term, $g_{nm} \equiv \sum_\alpha V_{nm}^\alpha e^{S_n}Q_\alpha e^{-S_m}$, perturbatively. The transition rate between states can then be written as
\begin{equation}
    k_{n\to m} (t) = \frac{1}{\hbar^2}\int_{-t}^{t} d\tau e^{i\frac{\varepsilon_n-\varepsilon_m}{\hbar}\tau} \left<g_{nm}(\tau)g_{mn}(0)\right>_{\rm eq},
\end{equation}
where $\left<\cdot\right>_{\rm eq}$ represents the average over the equilibrium nuclear degrees of freedom. The correlation function in the above equation can be expressed analytically as~\cite{izmaylov2011nonequilibrium}
\begin{align}
     \left<g_{nm}(\tau)g_{mn}(0)\right>_{\rm eq} = \left[h(\tau)^2 + l(\tau)\right]f_{\rm FC}(\tau)
\end{align}
where 
\begin{align}\label{eq:fgr_analytic}
\begin{split}
    h(t) &= \sum_\alpha \frac{V_{nm}^\alpha}{2\omega_\alpha^2}\left(V_{nn}^\alpha + V_{mm}^\alpha\right) + \frac{V_{nm}^\alpha}{2\omega_\alpha^2}\left(V_{nn}^\alpha - V_{mm}^\alpha\right) \left[(n_\alpha+1)e^{-i\omega_\alpha t} -n_\alpha e^{i\omega_\alpha t}\right]\\
    l(t) &= \sum_\alpha \frac{\hbar(V_{nm}^\alpha)^2}{2\omega_\alpha}\left[(n_\alpha+1)e^{-i\omega_\alpha t} + n_\alpha e^{i\omega_\alpha t}\right]\\
    f_{\rm FC}(t) &= \exp{\sum_\alpha -\frac{\left(V_{nn}^\alpha-V_{mm}^\alpha\right)^2}{2\hbar\omega_\alpha^3}(2n_\alpha+1) + \frac{\left(V_{nn}^\alpha-V_{mm}^\alpha\right)^2}{2\hbar\omega_\alpha^3}\left[(n_\alpha+1)e^{-i\omega_\alpha t} + n_\alpha e^{i\omega_\alpha t}\right]}. 
\end{split}
\end{align}
In Eq.~\eqref{eq:fgr_analytic}, Bose factor $n_\alpha = 1/(e^{\beta\hbar\omega_\alpha}-1)$, and the Franck-Condon prefector $f_{\rm FC}(t)$ is the same term appearing in the Marcus and Forster theories. 

The relaxation dynamics were then generated using a quantum master equation with rates computed using Eq.~\eqref{eq:fgr_analytic}:
\begin{align}
    \frac{dp_n(t)}{dt} &= \sum_{\substack{m \\ m\neq n}}p_m(t) k_{m\to n}(t) - \sum_{\substack{m \\ m\neq n}}p_n(t) k_{n\to m}(t).
\end{align}
or in the matrix form
\begin{equation}\label{eq:master}
    \frac{d\mathbf{p}(t)}{dt} = \mathbf{K}(t)\mathbf{p}(t),
\end{equation}
where
\begin{equation}
    K_{nm}(t) = k_{m\to n}(t) \hspace{0.2cm}(n\neq m), \hspace{0.5cm} K_{nn}(t) = -\sum_{\substack{m \\ m\neq n}}K_{mn}(t). 
\end{equation}
and $p_n(t)$ is the population of state $\ket{\psi_n}$. For the relaxation dynamics of CdSe NCs, we made an additional Markovian approximation where $k_{nm}(t)$ is replaced with its long-time limit $k(\infty)$. This is suitable for hot electron cooling with timescales much longer than the bath response time.

Fig.~3 in the main text plots the population $p_n(t)$ for electron relaxation among three 1P and 1S states. The initial population was placed in the highest energy 1P state.

\section{Radiative lifetimes}
We use the following expression to estimate the radiative lifetimes as $\tau_{\rm rad} = 1/k_{\rm {rad}}$:\cite{lakowicz1999introduction}
\begin{equation}
k_{\rm {rad}}=\frac{2 \omega_{\rm 1P\rightarrow 1S}^3}{3 \varepsilon_o h c^3} |\bra{1S}\bm{\mu}\ket{1P}|^2
\end{equation}
In the above, $\omega_{\rm 1P\rightarrow 1S} = (E_{\rm 1P} - E_{\rm 1S})/\hbar$, $c$ is the speed of light, $\varepsilon_0$ is the vacuum permittivity, and the transition dipole moment is computed using the $\ket{1P}$ and $\ket{1S}$ states.

\input{supplement.bbl}

%% file: main.bbl
\providecommand{\latin}[1]{#1}
\makeatletter
\providecommand{\doi}
  {\begingroup\let\do\@makeother\dospecials
  \catcode`\{=1 \catcode`\}=2 \doi@aux}
\providecommand{\doi@aux}[1]{\endgroup\texttt{#1}}
\makeatother
\providecommand*\mcitethebibliography{\thebibliography}
\csname @ifundefined\endcsname{endmcitethebibliography}
  {\let\endmcitethebibliography\endthebibliography}{}

%% file: supplement.bbl
\providecommand{\latin}[1]{#1}
\makeatletter
\providecommand{\doi}
  {\begingroup\let\do\@makeother\dospecials
  \catcode`\{=1 \catcode`\}=2 \doi@aux}
\providecommand{\doi@aux}[1]{\endgroup\texttt{#1}}
\makeatother
\providecommand*\mcitethebibliography{\thebibliography}
\csname @ifundefined\endcsname{endmcitethebibliography}
  {\let\endmcitethebibliography\endthebibliography}{}